\documentclass{article}

\usepackage{arxiv}

\usepackage[utf8]{inputenc}
\usepackage[T1]{fontenc}
\usepackage{amsfonts}
\usepackage{amsmath}
\usepackage{booktabs}
\usepackage{graphicx}
\usepackage{microtype}
\usepackage[authoryear,round]{natbib}
\bibpunct{(}{)}{;}{a}{}{, }
\usepackage{algorithm}
\usepackage{algpseudocode}
\usepackage{hyperref}
\usepackage{doi}
\usepackage[nameinlink]{cleveref}
\usepackage{tabularx}
\usepackage{orcidlink}

\graphicspath{{figures/}{assets/}}

\title{Beyond Executable Models: The Pufibara Agent Harness and the Modelica Agent Workflow Benchmark for Physical System Modeling}

\author{
  {\large\orcidlink{0009-0009-9685-2729}}\,
  \textbf{Wang Zizhe}\\
  \texttt{zizhe.wang@tu-dresden.de}
}
\date{}

\renewcommand{\shorttitle}{The Pufibara Agent Harness and the Modelica Agent Workflow Benchmark}

\hypersetup{
  colorlinks=true,
  linkcolor=blue,
  citecolor=blue,
  urlcolor=blue,
  pdftitle={Beyond Executable Models: The Pufibara Agent Harness and the Modelica Agent Workflow Benchmark for Physical System Modeling},
  pdfsubject={Artificial Intelligence (cs.AI)},
  pdfauthor={Wang Zizhe},
  pdfkeywords={agentic AI, physical system modeling, Modelica, simulation, benchmarks},
}

\begin{document}
\maketitle

\begin{abstract}
AI agents are increasingly used for simulation-driven engineering. Physical system modeling presents different requirements from general-purpose code generation in software engineering, because correctness depends not only on syntax and executability but also on physical consistency and scenario-dependent behavior. We study this challenge in Modelica, an equation-based modeling language in which a model may compile and simulate while still violating its intended physics or engineering requirements. Across successive revisions, an agent may lose track of requirements or rely on simulation evidence produced by an outdated candidate.

To address this challenge, we present Pufibara, an agent harness that maintains persistent engineering state across revisions, associates execution and simulation evidence with the candidate that produced it, and makes submission an explicit agent action. To evaluate end-to-end Modelica agent workflows, we also propose a source-grounded method for constructing realistic and independently evaluable tasks. We use this method to build the 232-task Modelica Agent Workflow Benchmark, spanning Model Repair, Model Generation, and Model Tuning. Each submitted candidate is scored by a benchmark-owned evaluator outside the agent loop.

We compare Pufibara and Claude Code as complete harnesses under two matched large language model (LLM) backends. With DeepSeek v4 Flash, Pufibara passes 202 tasks, compared with 185 for Claude Code. With Claude Sonnet 5, Pufibara passes 202 tasks, compared with 187 for Claude Code. Under the repository-reported token accounting, Pufibara records 76.4\%--82.5\% lower logical-token totals. Its sequential runtime is 6.1\%--58.4\% lower. These findings show that, even under matched LLM backends, complete agent harnesses can differ substantially in both task success and resource use for physical system modeling.
\end{abstract}

\begin{center}
\small
Agent Harness:
\href{https://github.com/wangzizhe/Pufibara}
{\texttt{https://github.com/wangzizhe/Pufibara}}

%\vspace{2pt}
Benchmark:
\href{https://github.com/wangzizhe/modelica-agent-workflow-benchmark}
{\texttt{https://github.com/wangzizhe/modelica-agent-workflow-benchmark}}
\end{center}

\section{Introduction}
\label{sec:introduction}

AI agents combine large language models (LLMs) with execution harnesses for tool use and iterative problem solving \citep{yao2023react}. An agent harness mediates the LLM's interaction with tools and its environment, maintains task state and execution feedback, and defines termination and submission semantics. These systems range from general-purpose agents such as Manus, Genspark, and OpenClaw to coding-oriented agents such as Claude Code, Codex, and DeepSeek Harness. Applying them to physical system modeling is promising, but executability alone does not establish the correctness of a physical system model.

We study this problem in Modelica, an open, equation-based, object-oriented language for modeling complex multi-domain cyber-physical systems \citep{fritzson1998modelica}. In many general-purpose imperative languages, an assignment specifies a direction of computation: the right-hand side is evaluated and stored in the variable on the left. In a Modelica equation section, an equality instead states a relation among quantities without specifying which variable must be solved for. The Modelica translator determines computational causality as it transforms the coupled equation system for numerical solution \citep{modelicaassociation2026specification}. Evaluating a generated model therefore requires more than checking syntax and executability. Its equations and simulated trajectories must also be consistent with the intended physics and scenario-dependent engineering requirements.

LLMs can generate plausible Modelica code, while AI agents extend this capability through iterative inspection, modification, and simulation. Yet neither plausible generation nor successful execution ensures that a model satisfies task-specific physical constraints and behavioral requirements. Over an iterative run, the agent may lose track of requirements, while simulation results may remain in context after the model that produced them has changed. The agent may therefore submit an executable model that does not satisfy the engineering brief.

Our goal is to evaluate AI agents as end-to-end Modelica problem-solving systems rather than evaluate only the models they generate. To our knowledge, no existing benchmark evaluates the full Modelica agent workflow, from interpreting an engineering brief through tool use and simulation-driven revision to explicit submission.

Constructing such a benchmark presents an additional challenge. Publicly accessible system-level Modelica models are scarce and concentrated in a small number of open libraries, whereas industrial models are generally proprietary. Public examples may already have appeared in LLM pretraining data. Evaluating them unchanged can therefore conflate task-solving ability with memorization or reproduction of previously seen artifacts. Conversely, arbitrarily authored synthetic tasks may lack realistic physical structure or trustworthy evaluation criteria.

To address these challenges, this paper makes three contributions:

\begin{itemize}
\setlength{\itemsep}{0.4em}
\item \textbf{Agent harness.}
We present Pufibara, an agent harness that preserves engineering requirements across model revisions and ties execution and simulation results to the candidates that produced them.

\item \textbf{Benchmark.}
We propose a source-grounded method for constructing realistic and independently evaluable tasks, and use it to build the Modelica Agent Workflow Benchmark for end-to-end agentic Modelica workflows.

\item \textbf{Evaluation.} 
We compare Pufibara with Claude Code across 232 tasks under two matched LLM backends. Pufibara achieves higher pass counts with lower reported logical-token use and runtime under both backends.
\end{itemize}

To our knowledge, this work introduces the first agent harness specifically designed for end-to-end Modelica workflows and the first benchmark for evaluating those workflows. Under matched LLM backends, the results show that harness design is associated with substantial differences in task success and resource use.
\section{Related Work}
\label{sec:related-work}

\subsection{Agent Harnesses and Execution Feedback}
\label{sec:related-agent-harnesses}

ReAct \citep{yao2023react} interleaves LLM reasoning with actions and observations from an external environment. CodeAct \citep{wang2024codeact} uses executable code as its action representation, allowing agents to execute code and revise their actions in response to execution feedback. AI coding harnesses such as Claude Code\footnote{\url{https://claude.com/claude-code}}, Codex\footnote{\url{https://openai.com/codex}}, and DeepSeek Harness\footnote{\url{https://deepseek.com/harness}} operationalize these interaction patterns by connecting LLMs to repository navigation, file editing, command execution, and iterative feedback. Their interfaces, context management, tool protocols, and completion semantics shape how the underlying models behave. Together, these systems establish the action-observation loop as a central pattern for connecting an LLM to its environment. 

In software engineering, SWE-bench \citep{jimenez2024swebench} introduces repository-level tasks drawn from real GitHub issues and assesses LLM-generated patches using executable tests. SWE-agent \citep{yang2024sweagent} extends this setting to interactive coding agents through an Agent-Computer Interface for repository navigation, code editing, and test execution. OpenHands \citep{wang2025openhands} further generalizes this architecture into a platform in which agents interact with repositories, command-line tools, code execution environments, and web resources. Agentless \citep{xia2025agentless} provides a contrasting design based on fixed localization, repair, and validation stages rather than an autonomous tool-use loop. Together, these systems show that capability depends not only on the underlying LLM, but also on the interface, action space, execution environment, and workflow through which the agent operates.

In these settings, repository state and executable tests provide the primary feedback for revision and the primary criterion for accepting a patch. Physical system modeling additionally requires an agent to evaluate continuous, scenario-dependent simulation trajectories against physical constraints and behavioral requirements. Extending such workflows to this domain therefore requires engineering obligations to remain explicit across iterations and simulation evidence to be associated with the specific candidate that produced it.

\subsection{AI Agents for Simulation-Driven Engineering}
\label{sec:related-simulation-agents}

Recent work has applied AI agents to simulation-driven engineering. \citet{moeltner2026creation} generate Python multibody simulation models and study self-validation by agents using predefined validation procedures and expert-created reference models. SimuGen \citep{ren2025simugen} coordinates specialized agents to construct, execute, and debug Simulink models from block diagrams and domain knowledge. SimuAgent \citep{liang2026simuagent} combines a plan-execute architecture with simulation feedback and evaluates a trained Simulink modeling agent on the SimuBench task collection. ASWE-Bench \citep{abdalla2026aswebench} evaluates a multi-agent workflow for automotive Simulink model construction, test generation, compliance checking, and iterative refinement across 38 software requirements. 

FEABench \citep{mudur2025feabench} evaluates LLMs and agents that operate COMSOL Multiphysics through its API, inspect solver outputs, and iteratively improve solutions to multiphysics problems. MCP-SIM \citep{park2026mcpsim} uses persistent shared memory and specialized agents to construct, execute, diagnose, and revise finite element simulations from natural language requests. Dyad \citep{lima2026dyad} integrates a specialized multi-agent workflow with an acausal equation-based modeling language and evaluates it on process system modeling, validation, and control tasks. Agent-in-the-Loop \citep{bjorkskog2026agent} evaluates an LLM agent using Functional Mock-up Units for system identification and iterative proportional-integral controller tuning.

\citet{agrawal2024coupled} separately emphasize traceability from system requirements to simulation scenarios and the resulting test evidence in cyber-physical systems. PowerAgentBench-SS \citep{mylonas2026poweragentbench} provides a closely related evaluation setting. It evaluates tool-using agents that inspect power system cases, invoke simulators, validate mitigations, and submit an auditable evidence trail. A hidden evaluator then independently recomputes physical validity. PowerAgentBench-Dyn \citep{zhang2026poweragentbenchdyn} extends agent evaluation to dynamic model review and simulation-driven contingency analysis. These benchmarks study operational analysis and mitigation over fixed power system cases, whereas the present work evaluates construction, repair, and tuning of equation-based model artifacts across candidate revisions. Together, these works highlight the value of simulation feedback, automated validation, requirement-linked testing, and independent physical evaluation.

These systems operate in complementary settings, including finite element analysis, multibody simulation, Simulink, process systems engineering, and steady-state power system analysis. End-to-end evaluation of a common agent workflow for constructing and revising equation-based physical system models remains underexplored.

\subsection{Modelica Generation and Benchmarking}
\label{sec:related-modelica}

Recent Modelica research has primarily focused on code generation and benchmark datasets. Text2Model \citep{rupprecht2025text2model} generates dynamic chemical reactor models from textual descriptions and evaluates their syntactic and semantic accuracy. ModiGen \citep{xiang2025modigen} introduces datasets for Modelica component and test case generation and combines supervised fine-tuning, graph retrieval, and feedback optimization. \citet{wan2025modelica} combine structured prompts, library-aware grounding, automated OpenModelica compilation, and human review to generate control modules for the Building Modelica Library. \citet{stuermer2026fluid} present a benchmark for fluid systems that compares LLMs and prompting strategies for translating graph representations into Water Network Tool for Resilience (WNTR) and Modelica code. The benchmark evaluates both software quality and simulation fidelity. These studies expose important limitations in Modelica generation, but their evaluations focus on generated artifacts or task-specific generation workflows rather than complete agent workflows involving iterative tool use, simulation-driven revision, and explicit submission.

ModBench \citep{sadrnezhaad2026modbench} provides a pipeline for mining Modelica repositories and constructing a large, traceable dataset of class snapshots. The dataset supports research on model evolution, compiler testing, repair, and generation, but does not define an end-to-end agent evaluation workflow. 
%The Modelica Association has also announced an ongoing community effort to assemble automatically graded AI tasks covering model construction, simulation diagnosis, and parameter tuning\footnote{\url{https://modelica.org/newsletter/2026-01}}.

Recent developments in the Modelica software ecosystem provide agent-facing access to model editing, checking, and simulation through the Model Context Protocol (MCP). These include new Modelica-focused tools such as ODE Plus\footnote{\url{https://www.orthogonal.dev}} and Modex\footnote{\url{https://modexai.io}}. Established Modelica platforms are also adding related agent capabilities. These integrations extend the tools available to general-purpose agents, but do not themselves define the agent loop, persistent engineering state, or submission protocol of an agent harness designed specifically for Modelica.

Together, these efforts cover generated artifacts, traceable Modelica datasets, automatically graded tasks, and agent-accessible tooling. To our knowledge, however, prior work has not introduced an agent harness specifically designed for end-to-end Modelica workflows or a benchmark for evaluating those workflows.
\section{Engineering Invariants and Candidate-Bound Evidence}
\label{sec:method}

This section defines a method for maintaining engineering requirements and candidate-specific evidence across iterative simulation-driven engineering tasks. It represents requirements as persistent obligations and binds them to observables, scenarios, and simulation evidence. The abstraction is independent of any particular physical domain, while the engineering propositions, observables, scenarios, and evidence conditions remain task-specific. The agent uses this state to decide whether to collect further evidence, revise the candidate, or submit it, while an independent evaluator determines the official task outcome.

\subsection{Engineering Invariants as Persistent Obligations}
\label{sec:invariant-ledger}

\citet{sargent2013verification} distinguishes verification of a model's implementation from validation for its intended use. Accordingly, a physical system model may be syntactically valid, accepted by a compiler, and successfully simulated while still violating its engineering requirements. Compilation establishes that the model can be translated, and simulation establishes that it can execute under a particular configuration. Neither establishes that the resulting behavior represents the intended physical system.

\citet{hu2026memoryagentbench} report limitations in accurate retrieval, long-range understanding, and selective forgetting during incremental multi-turn interactions. These limitations become consequential in long-horizon engineering workflows. As an agent repeatedly inspects, modifies, and simulates a model, its working representation of the original requirements may drift. Assumptions may be forgotten, evidence from an earlier candidate may be reused after the model changes, and successful execution may be mistaken for satisfying the task requirements. These failure modes show why execution alone is insufficient as a success criterion. End-to-end physical system modeling requires a persistent representation of what must remain true, how those requirements can be observed, and what evidence is needed before submission.

To provide this persistent representation, we use \emph{engineering invariant} as an operational term for a task-level obligation that remains in force across modeling iterations and must be addressed before submission. Here, invariant refers to the persistence of the obligation, not necessarily to a quantity that remains constant over simulation time. Engineering invariants can take several forms. Physical invariants express relationships such as conservation, sign, boundedness, or consistency between physical quantities. Engineering constraints specify required interfaces, structures, parameter bounds, or operating limits. Behavioral requirements describe how the modeled system should respond under a defined scenario. For example, a Modelica task may require conservation across connected components, specified connector interfaces and parameter bounds, and settling within a target range after an input change.

Systems such as A-MEM \citep{xu2025amem} organize evolving information outside transient conversational context through structured indexing, linking, and updating. An \emph{engineering invariant ledger} serves a more specific role: it organizes persistent state around the obligations that guide revision and submission. For physical system modeling, an agent authors the ledger from the engineering brief and maintains it across modeling iterations. Its purpose is to prevent required constraints and behaviors from being silently displaced by more recent execution feedback.

An invariant specification can be represented as

\begin{equation}
    i_k =
    \left\langle
        p_k,\,
        O_k,\,
        S_k,\,
        \phi_k
    \right\rangle ,
\end{equation}

where \(p_k\) is the engineering proposition, \(O_k\) is the set of relevant observables, \(S_k\) is the set of simulation scenarios under which it should be examined, and \(\phi_k\) describes the condition by which evidence is judged. This condition may be quantitative, such as a bound or tolerance, or qualitative, such as an expected response pattern.

For the current candidate model \(c\), the ledger records a working status \(\sigma_k(c)\) for each invariant. The status may be open, supported, violated, or inconclusive. It must be reconsidered whenever the candidate changes because the current status may be based on evidence from an earlier model. The ledger is procedural rather than authoritative: it makes the agent's interpretation explicit, persistent, and auditable and identifies which engineering obligations still require attention, but it does not determine the official correctness of the submission.

\subsection{Observable and Scenario Bindings for Candidate-Bound Evidence}
\label{sec:evidence-binding}

An invariant cannot guide an agent merely by remaining in natural language. It must be connected to observable model behavior. We call this connection an \emph{observable binding}. The binding identifies observables whose simulated behavior can provide evidence about the proposition. An observable may be a directly exposed model variable or a derived trajectory-level quantity, such as an extremum, settling time, integral, rate of change, or relationship among signals.

The monitoring framework of \citet{maler2004monitoring} evaluates temporal properties over continuous signals. \citet{agrawal2024coupled} likewise connect system requirements to simulation scenarios and resulting test evidence in cyber-physical system testing. Accordingly, observable bindings are paired with simulation scenarios. A scenario specifies the conditions under which the proposition is examined, including relevant parameters, inputs, disturbances, initial conditions, and time intervals. For example, a requirement that pressure return to a target range after a load change must be bound to a pressure observable, a load-change scenario, an evaluation interval, and conditions describing the acceptable response.

\citet{villamar2025metadata} emphasize that simulation results must remain connected to metadata describing the model, configuration, inputs, and execution that produced them. To preserve the relevant provenance, we represent each result in an evidence record as

\begin{equation}
    e_j =
    \left\langle
        i_k,\,
        c,\,
        s_j,\,
        O_j,\,
        Y_j,\,
        a_j
    \right\rangle ,
\end{equation}

where \(i_k\) identifies the invariant addressed by the evidence, \(c\) identifies the candidate model, \(s_j \in S_k\) identifies the simulation scenario, \(O_j \subseteq O_k\) identifies the inspected observables, \(Y_j\) contains the resulting trajectories or derived measurements, and \(a_j\) records the agent's adjudication of the evidence against \(\phi_k\).

Candidate identity is essential. Evidence from an earlier candidate remains part of the audit history, but it cannot support the readiness of a revised candidate. The relevant invariants must therefore be reconsidered using evidence generated from the revised candidate.

Evidence adjudication classifies a result as supporting, violating, or remaining inconclusive with respect to an invariant. Successful simulation alone is not supporting evidence. The resulting observations must be interpreted against the engineering proposition that motivated the simulation.

\subsection{Revision, Readiness, and Explicit Submission}
\label{sec:submission-readiness}

The resulting workflow can be summarized as

\begin{equation}
\begin{split}
    \text{engineering brief}
    &\rightarrow \text{invariant ledger}
    \rightarrow \text{observable and scenario bindings} \\
    &\rightarrow \text{candidate model}
    \rightarrow \text{targeted simulation}
    \rightarrow \text{evidence adjudication} \\
    &\rightarrow
    \{\text{further evidence},\,\text{revision},\,\text{submission}\}.
\end{split}
\end{equation}

The agent first translates the engineering brief into a ledger of explicit obligations. It then constructs or modifies a candidate, selects simulations that address unresolved invariants, and adjudicates the resulting evidence. Open or uncovered invariants trigger targeted evidence collection. Violated invariants trigger model revision. Inconclusive evidence triggers either additional observation or a more informative simulation scenario. Candidate modification, in turn, requires the relevant evidence to be refreshed.

Let \(\mathcal{I}_{\mathrm{req}}\) denote the required invariants and let \(\mathcal{E}_c\) denote the evidence bound to candidate \(c\). The workflow considers a candidate ready for submission when

\begin{equation}
    \operatorname{ready}(c)
    \Longleftrightarrow
    \forall i \in \mathcal{I}_{\mathrm{req}},
    \operatorname{covered}(i,\mathcal{E}_c)
    \land
    \operatorname{supported}(i,\mathcal{E}_c).
\end{equation}

Coverage requires the evidence for candidate \(c\) to include the observables and scenarios specified by the invariant. Support requires the evidence associated with each required scenario to satisfy the corresponding condition, with no required scenario left violated or inconclusive. The readiness predicate is therefore specific to both the candidate and the engineering brief.

Submission is an explicit action rather than an implicit consequence of producing or simulating a model. This separates the most recently edited candidate from the candidate that the agent intentionally selects for submission. The readiness rule guides the agent's decision to submit, while an independent evaluator remains responsible for determining the official task outcome.

This formulation extends the conventional read-edit-test loop by binding simulation evidence to persistent engineering obligations, observables, scenarios, and candidate identity. The agent adjudicates this evidence before deciding whether to collect further evidence, revise the candidate, or submit it.
\section{The Pufibara Agent Harness}
\label{sec:pufibara}

This section instantiates the method in \Cref{sec:method} as Pufibara, an agent harness for simulation-driven engineering. Each run begins with a task package containing an engineering brief, optional initial artifacts, and workspace context. Pufibara mediates the agent's access to these inputs, persistent engineering state, workspace tools, model checking, and simulation. The run ends with an explicit submission or, if a termination condition is reached first, without one. Throughout the run, Pufibara leaves engineering decisions to the agent while preserving the state and provenance needed to associate actions and evidence with candidate revisions. This separation between agent decision-making and harness-managed execution also provides a clear boundary for independent evaluation.

\subsection{Harness Architecture and Workflow Profiles}
\label{sec:pufibara-architecture}

\begin{figure}[htbp]
    \centering
    \includegraphics[width=\linewidth]{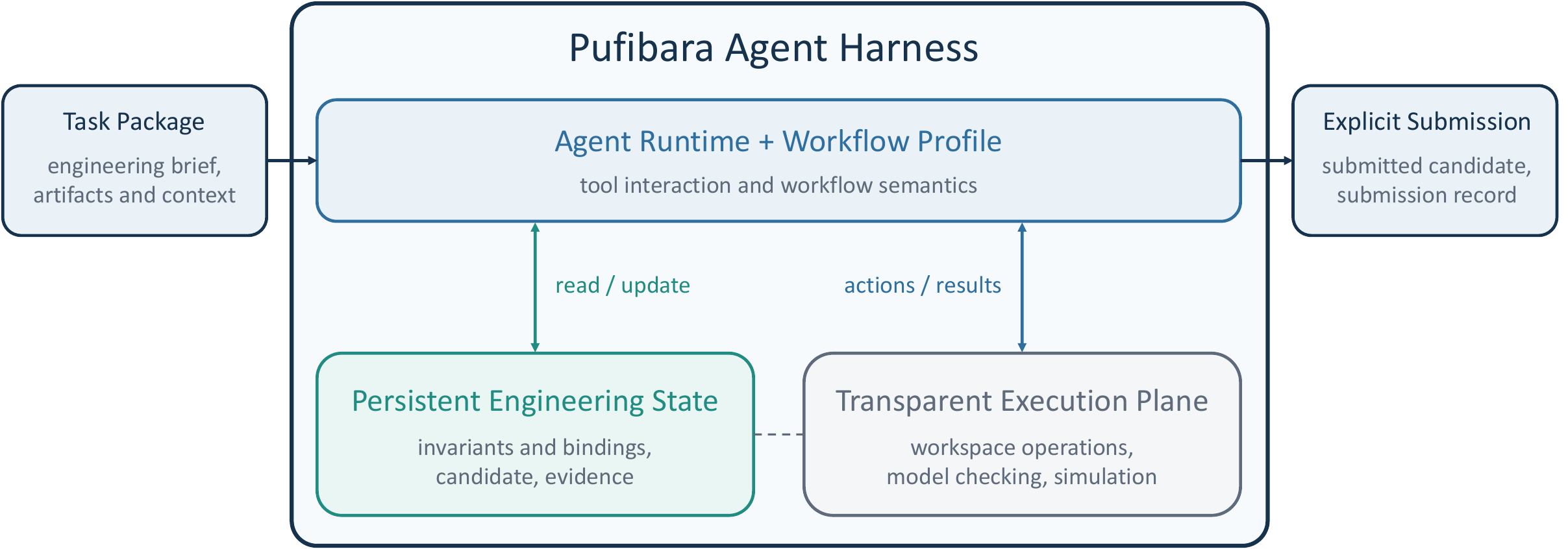}
    \caption{Architecture and run boundary of the Pufibara harness. Solid internal arrows show the runtime reading or updating persistent state and exchanging actions and results with the execution plane. The dashed line indicates that every execution result is associated with the candidate that produced it.}
    \label{fig:pufibara-architecture}
\end{figure}

As shown in \Cref{fig:pufibara-architecture}, Pufibara contains three internal components. The \emph{Agent Runtime}, together with a \emph{Workflow Profile}, mediates iterative interaction and defines workflow semantics. \emph{Persistent Engineering State} preserves task context, candidate identity, execution records, and candidate-bound evidence across agent turns. The \emph{Transparent Execution Plane} performs workspace operations, model checking, and simulation requested by the agent.

The runtime mediates the iterative exchange between the LLM and its environment. On each turn, it exposes the relevant task, workflow profile, persistent state, and available observations to the agent. The workflow profile specializes this exchange by defining the task representation, permitted actions, candidate representation, and explicit submission contract. It is therefore an execution protocol rather than merely a prompt template. The agent remains responsible for interpreting the task, constructing or modifying a candidate, selecting actions and simulations, adjudicating results, and deciding when to submit. The harness executes requested actions and records their outcomes and provenance, but it does not independently choose engineering actions or submit a candidate on the agent's behalf.

The three workflow profiles instantiate different candidate and submission semantics. Model Repair treats the supplied faulty model as the initial candidate and submits a repaired artifact set. Model Generation may begin without a complete candidate, permits the agent to construct the required model and package artifacts, and submits the resulting artifact set. Model Tuning keeps the model structure fixed, treats an admissible parameter assignment as the candidate, and submits parameter values. Despite these differences, all profiles share the same boundaries for transparent execution and explicit submission, while each profile defines its candidate representation and workflow-specific state.

\subsection{Persistent Engineering State and Transparent Execution}
\label{sec:pufibara-state}

\paragraph{Persistent Engineering State.}
Long-horizon modeling requires state beyond the transient conversation of a single agent turn. Pufibara therefore maintains a persistent run record containing the task and profile context, current candidate identity, workflow-specific engineering state, execution and simulation records tied to candidates, candidate transitions, prior actions, and any explicit submission. The runtime exposes the relevant part of this state on each turn rather than reconstructing it from the most recent tool result.

Within this state, the engineering invariant ledger can represent obligations that require explicit behavioral adjudication. Maintained by the agent, it records the physical invariants, engineering constraints, and behavioral requirements being tracked, together with their observable and scenario bindings, candidate-relative status, and references to adjudicated evidence. The ledger remains a working engineering representation rather than an official acceptance oracle.

Candidate identity is defined by the workflow. For Repair and Generation, it identifies the complete evaluation-relevant artifact state. For Tuning, it identifies the parameter assignment together with the frozen model and configuration to which that assignment applies. Any change to evaluation-relevant submitted content produces a distinct candidate identity. Earlier candidates and their records remain in the run history, while readiness for the current candidate considers only evidence bound to its identity. For example, if a simulation of candidate \(c_1\) supports a settling-time obligation, an artifact change that produces \(c_2\) preserves that result in the audit history but reopens the obligation for \(c_2\) until candidate-specific evidence is collected.

\paragraph{Transparent Execution Plane.}
The execution plane performs actions selected by the agent and permitted by the workflow profile. These include inspecting and modifying workspace artifacts, checking Modelica models, and running simulations. Execution is transparent in the sense that the requested action, target candidate, resulting observation, and produced artifacts are returned to the agent and recorded in the persistent state. The execution plane does not silently modify a candidate or decide the next engineering action.

A change to candidate-defining content creates a recorded transition from the source candidate to the resulting candidate. Model checking results remain associated with the candidate that was checked. A simulation result is additionally recorded with its scenario, inspected observables, configuration, and resulting trajectories or derived measurements. This provenance prevents a result produced from one workspace state from being silently treated as a result for a later state at the same path.

A recorded simulation result is not yet adjudicated evidence. The agent must relate it to the corresponding engineering obligation and determine whether it supports, violates, or fails to resolve that proposition. Pufibara then preserves the adjudication and its candidate binding. It does not determine whether the agent's physical interpretation is correct.

\subsection{Explicit Submission and the Evaluation Boundary}
\label{sec:pufibara-evaluation}

Submission is a distinct agent action. A candidate is not submitted merely because it compiles, simulates, or is the most recently modified artifact. The submission record freezes the current candidate identity and the exact artifact set or parameter assignment selected by the agent. The runtime does not promote the last runnable model, substitute an earlier candidate, or infer submission from tool success.

The readiness rule in \Cref{sec:submission-readiness} governs obligations represented by the active workflow profile. Before submission, each applicable obligation should be covered and supported by evidence bound to the current candidate. Pufibara exposes the state needed to apply this rule and records the resulting decision, but it neither certifies engineering correctness nor submits automatically. A run that reaches its termination condition without an explicit submission is recorded as a failure to submit.

After explicit submission, the agent's revision loop terminates. A benchmark-owned evaluator then applies a task-specific acceptance contract to the exact submitted artifact or parameter assignment. The agent's working state, simulations, and adjudications do not define this contract, and the resulting verdict is not returned as another opportunity for revision. The construction and validation of these contracts belong to the Modelica Agent Workflow Benchmark described in \Cref{sec:benchmark}.

This separation creates two distinct judgments. Agent-side readiness determines whether the agent considers a candidate sufficiently supported for submission. Benchmark-side acceptance determines the official outcome by applying a frozen contract to the exact submission. The evaluator does not use the agent's ledger or self-adjudication as acceptance criteria, so agent-authored validation cannot redefine the official success criterion.

With the state, execution, and submission boundaries defined, \Cref{alg:pufibara-loop} summarizes the complete run protocol. Here, \textit{state} is the persistent run record, and \textit{candidate} is the current model artifact set or parameter assignment. The value \textit{none} means that the run ended without an explicit submission.

\begin{algorithm}[htbp]
\caption{Pufibara agent loop for persistent state, transparent execution, and explicit submission.}
\label{alg:pufibara-loop}
\begin{algorithmic}[1]
\Require Task package \textit{task} and workflow profile \textit{profile}
\State Initialize persistent \textit{state}
\State Identify the initial \textit{candidate}, if one exists
\While{the run remains active}
    \State Expose \textit{task}, \textit{profile}, \textit{state}, and \textit{candidate} to the agent
    \State $\mathit{action} \gets$ next action selected by the agent
    \If{$\mathit{action}$ explicitly submits the current candidate}
        \State Freeze the exact candidate and submitted content
        \State \Return the frozen submission
    \ElsIf{$\mathit{action}$ updates persistent engineering state}
        \State Persist the update for the current candidate
    \Else
        \State Execute the action transparently
        \If{$\mathit{action}$ changes candidate-defining content}
            \State Identify and record the resulting candidate
            \State Make it the current candidate
        \Else
            \State Record the result with the current candidate
        \EndIf
        \State Return the resulting observation or artifact to the agent
    \EndIf
\EndWhile
\State \Return \textit{none}
\end{algorithmic}
\end{algorithm}

The persistent run record links the task identity, agent actions, candidate transitions, execution results, evidence adjudications, and submission. The benchmark can therefore associate its independent outcome with the exact submission without entering the harness's engineering decision loop. With the harness protocol defined, the next section describes the task packages and independent evaluation contracts used to evaluate complete Modelica agent workflows.

\section{Modelica Agent Workflow Benchmark}
\label{sec:benchmark}

Evaluating end-to-end Modelica agent workflows requires tasks that are realistic, novel, and independently evaluable. Publicly accessible system-level Modelica models are scarce, concentrated in a small number of open libraries, and may already have appeared in LLM training data. Evaluating these models unchanged can therefore turn the benchmark into a test of memorization rather than problem solving. Creating arbitrary synthetic tasks avoids direct reuse but may produce unrealistic physical systems and unreliable evaluation criteria. Even when a suitable source model is available, the model alone does not define an engineering brief, executable root, dependency context, or evaluator.

To balance these requirements, we propose \emph{source-grounded task synthesis}. Each task begins with an executable clean reference, from which a new fault, engineering brief, tuning target, or evaluation contract is constructed. The reference provides realistic physical structure, but it is not itself the answer to the resulting task.

Using this method, we build the Modelica Agent Workflow Benchmark for evaluating agents as complete iterative systems rather than scoring one-shot LLM responses. Each run begins with an agent-visible task package and ends when the agent explicitly submits a candidate or reaches the execution limit. In the reported experiments, the operative limit is a 900-second wall-clock timeout, while the turn and token limits serve as high guardrails. A benchmark-owned evaluator scores only the exact submitted output. A run that ends without submission is recorded as a failure to submit.

\subsection{Source-Grounded Task Construction}
\label{sec:benchmark-construction}

All 232 tasks are grounded in executable reference models that pass their applicable checks. Of these, 140 are directly derived from public models, 15 compose components from public Modelica libraries, including the Modelica Standard Library \citep{modelicaassociation2026msl}, into new systems, and 77 use internally authored references. Internally authored references may use public library components, but their top-level systems are not copies of published example models.

Task derivation is workflow specific. Repair tasks introduce controlled faults into reference models. Generation tasks derive new engineering briefs, interfaces, observables, and behavioral contracts from reference systems. Tuning tasks preserve a fixed model while defining bounded parameter spaces, target responses, and evaluation scenarios. In each case, the resulting task differs from its source model even when the underlying library components are publicly known.

Each task fixes an exact top-level model, dependency context, and simulation configuration. At evaluation time, the benchmark reconstructs the corresponding package workspace and evaluates the submitted candidate against that fixed root. This preserves the package and library structure in which the agent operates rather than flattening the package into a single model file.

Task construction was validated separately for each workflow. Each faulty Repair input had to fail at its intended stage before being included in the benchmark: 123 failed during model checking and nine during simulation. For every Generation and Tuning task, the evaluator was tested against both a known-valid answer and deliberately incorrect but executable variants. The valid answer had to pass, while the off-target variants had to fail the task-specific behavior check. These checks confirm that the evaluator assesses the intended engineering behavior rather than merely rejecting malformed submissions. Task packages, evaluators, dependencies, simulation settings, and scoring rules were fixed before agent evaluation.

New faults, briefs, targets, and private evaluation contracts prevent an unchanged public source model from directly solving the task. This design reduces the risk of evaluating memorization, but it cannot guarantee that an underlying model or library component was absent from LLM pretraining data.

\subsection{Benchmark Scope and Workflow Families}
\label{sec:benchmark-scope}

The benchmark contains 232 tasks: 132 Model Repair tasks, 50 Model Generation tasks, and 50 Model Tuning tasks. Difficulty is assigned using empirical agent performance and workflow complexity rather than source-code length alone. \Cref{tab:benchmark-workflows} summarizes the workflow objectives, evaluation criteria, and difficulty distribution.

\begin{table}[htbp]
\centering
\small
\renewcommand{\arraystretch}{1.2}
\caption{Workflow families and difficulty distribution. E/M/H denote easy, medium, and hard.}
\label{tab:benchmark-workflows}

\begin{tabularx}{\linewidth}{
    @{}
    l
    >{\centering\arraybackslash}p{0.18\linewidth}
    >{\raggedright\arraybackslash}p{0.24\linewidth}
    >{\raggedright\arraybackslash}X
    @{}
}
\toprule
Workflow & Tasks (E/M/H) & Agent objective & Independent evaluation \\
\midrule
Repair
& 132 (21/56/55)
& Repair a faulty model to satisfy the task specification
& Interface preservation, model checking, and successful simulation \\

Generation
& 50 (2/10/38)
& Construct a model from an engineering brief
& Structural requirements, model checking, simulation, and behavioral contracts \\

Tuning
& 50 (4/24/22)
& Tune permitted parameters toward behavioral targets
& Parameter validity, successful simulation, and scenario-dependent response metrics \\
\midrule
Total
& 232 (27/90/115)
&
& \\
\bottomrule
\end{tabularx}
\end{table}

The tasks span electrical, magnetic, electromechanical, mechanical, thermal-fluid, building, process, control, and hybrid dynamics. \Cref{tab:benchmark-domains} reports one descriptive primary domain for each task.

\begin{table}[htbp]
\centering
\small
\renewcommand{\arraystretch}{1.2}
\caption{Task distribution by primary domain. Cross-domain tasks are counted under their dominant domain.}
\label{tab:benchmark-domains}

\begin{tabular*}{\linewidth}{
    @{\extracolsep{\fill}}
    lcccc
    @{}
}
\toprule
Primary domain & Repair (132) & Generation (50) & Tuning (50) & Total (232) \\
\midrule
Electrical, magnetic, and electromechanical & 57 & 11 & 23 & 91 \\
Thermal-fluid, building, and process systems & 57 & 20 & 17 & 94 \\
Mechanical and multibody & 9 & 8 & 10 & 27 \\
Control, signal, and hybrid dynamics & 9 & 11 & 0 & 20 \\
\bottomrule
\end{tabular*}
\end{table}

\subsection{Independent Evaluation Protocol}
\label{sec:benchmark-evaluation}

Each task has two parts. The \emph{agent-visible package} contains the engineering brief, permitted artifacts, public constraints, and required output format. The \emph{benchmark-owned contract} contains the evaluator configuration and, where applicable, private response metrics, scenarios, or reference artifacts. The contract is authored and frozen during benchmark construction and excluded from the agent workspace.

For task $t$ and submission $x$, the official decision can be summarized as

\begin{equation}
    \operatorname{PASS}(t,x)
    = I_t(x) \land M_t(x) \land C_t(x)
    \land S_t(x) \land B_t(x),
\end{equation}

where $I_t$ checks the submission interface, $M_t$ checks task-specific model or parameter constraints, $C_t$ checks successful Modelica model checking, $S_t$ checks successful simulation, and $B_t$ checks task-specific behavioral or response conditions. A task passes only if every applicable gate passes. The behavioral gate applies to all 50 Generation and all 50 Tuning tasks. Repair checks the required interface, model checking, and simulation outcome without a separate behavioral gate.

Each task uses a fresh agent session and isolated workspace, with no conversation, candidate, or engineering state reused across tasks. The evaluator runs only after explicit submission and receives the exact submitted artifact or parameter set. Its verdict is not returned for further revision. The agent may choose its own working validation procedure, but it cannot alter the official acceptance conditions.

Independence alone does not make an evaluator correct. An incorrect observable, scenario, tolerance, simulation grid, or sign convention can create false acceptance or rejection. Evaluator validation is therefore part of benchmark construction rather than an assumption that a runnable scoring script is a ground-truth oracle \citep{barr2015oracle}.

Known-valid reference artifacts and target configurations are evaluated through the same submission path as agent outputs and must pass. Executable but deliberately off-target variants must fail the applicable behavioral checks. The private contract is also reviewed against the agent-visible brief to ensure that it does not introduce contradictory requirements or demand a single literal implementation. These checks are repeated whenever an evaluator or execution dependency changes.

These checks reduce but do not eliminate evaluator risk. Finite scenarios cannot establish universal physical correctness. A PASS means that the submitted artifact satisfies the frozen, scenario-covered engineering contract. It is not formal verification or a guarantee over every operating condition.

\section{Evaluation}
\label{sec:evaluation}

This section compares the Pufibara and Claude Code harnesses on the full 232-task Modelica Agent Workflow Benchmark under two matched LLM backend conditions. It evaluates task success, resource use, and whether observed differences extend beyond model executability.

\subsection{Evaluation Scope and Design}
\label{sec:evaluation-design}

The evaluation considers three dimensions. Task success is measured by pass counts across the full benchmark and its three workflows. Resource use is measured by logical-token use and sequential wall-clock runtime. Correctness beyond executability is examined by identifying submissions that pass model checking and simulation but fail the independent behavioral contract.

The evaluation compares Pufibara with Claude Code under two LLM backend conditions, DeepSeek v4 Flash\footnote{DeepSeek v4 Flash refers to DeepSeek-V4-Flash-0731, accessed through the DeepSeek API alias \texttt{deepseek-v4-flash}.} and Claude Sonnet 5\footnote{Claude Sonnet 5 refers to the Anthropic API model identifier \texttt{claude-sonnet-5}, used with explicit medium effort.}. Within each combination of backend and workflow, both harnesses use the same underlying LLM and are evaluated on the same task set, Modelica environment, and benchmark-owned evaluator.

The comparison preserves each harness's native workflow semantics, context management, tool interaction, state representation, and submission protocol. It therefore compares the harnesses end to end rather than isolating individual mechanisms. Each task is evaluated once for each harness-backend combination, so the results characterize system-level differences under the reported conditions. They do not establish statistical significance or universal superiority.

\subsection{Experimental Setup}
\label{sec:experimental-setup}

\Cref{tab:experimental-setup} summarizes the experimental design. Exact software versions, execution limits, failure-handling rules, and token-field mappings are provided in \Cref{app:agent-configurations}.

\begin{table}[htbp]
\centering
\small
\renewcommand{\arraystretch}{1.15}
\caption{Evaluation setup.}
\label{tab:experimental-setup}
\begin{tabularx}{\linewidth}{
    @{}
    >{\raggedright\arraybackslash}p{0.25\linewidth}
    >{\raggedright\arraybackslash}X
    @{}
}
\toprule
Dimension & Configuration \\
\midrule
Systems
& Pufibara and Claude Code under DeepSeek v4 Flash and Claude Sonnet 5 \\

Matched conditions
& Same backend, tasks, Modelica environment, and evaluator \\

Execution
& One fresh isolated run per task, executed sequentially \\

Measures
& Task pass count, logical-token use, and sequential wall-clock runtime \\
\bottomrule
\end{tabularx}
\end{table}

Both harnesses receive only the agent-visible task package. DeepSeek v4 Flash is accessed through the official DeepSeek API, with its Anthropic-compatible endpoint used to expose the same backend to Claude Code. Claude Sonnet 5 is accessed through the official Anthropic API. Neither harness can access the private evaluator during the agent loop.

The primary effectiveness metric is the task pass count. A submission passes only if every applicable gate in \Cref{sec:benchmark-evaluation} passes. Results are reported by workflow and backend without pooling the two backends. Resource use is described using logical-token use and sequential task wall time. Logical-token use measures the total model-facing token volume accumulated across all LLM calls. For each call, it includes uncached input, cache-creation input, cache-read input, and output tokens:
\[
T_{\mathrm{logical}}
= T_{\mathrm{input}}
+ T_{\mathrm{cache\ creation}}
+ T_{\mathrm{cache\ read}}
+ T_{\mathrm{output}}.
\]
Cached input is counted because it remains part of the model-facing context, even when the provider bills it at a discounted rate. Logical-token use therefore characterizes the model interaction generated by a harness rather than its monetary API cost.

All reported resource reductions use Claude Code as the baseline within the same backend and workflow. A token reduction of 20\% means that Pufibara used 20\% fewer logical tokens than Claude Code, and runtime reductions are interpreted in the same way. Runtime totals sum valid task wall times under sequential execution. Logical-token totals are normalized operational measurements rather than provider billing totals.

\subsection{Benchmark Results}
\label{sec:evaluation-results}

\Cref{tab:benchmark-results} reports the benchmark results and relative differences under both LLM backends. With DeepSeek v4 Flash, Pufibara passes 202 tasks and Claude Code passes 185. With Claude Sonnet 5, the corresponding totals are 202 and 187. Pufibara has the higher observed pass count in all six combinations of backend and workflow.

\begin{table}[htbp]
\centering
\small
\renewcommand{\arraystretch}{1.15}
\caption{Benchmark results and relative differences between Pufibara and Claude Code. Token use is shown in millions. Pass-rate gains are reported in percentage points (pp), and token and runtime reductions are computed from unrounded repository totals.}
\label{tab:benchmark-results}
\begin{tabular*}{\linewidth}{
    @{\extracolsep{\fill}}
    llrrr
    @{}
}
\toprule
\multicolumn{5}{@{}l}{\textit{(a) DeepSeek v4 Flash}} \\
\midrule
Workflow & Harness & Passed & Tokens (M) & Runtime (h) \\
\midrule
Repair     & Pufibara   & 130/132 & 39.8  & 4.07 \\
           & Claude Code & 124/132 & 227.0 & 9.78 \\
\addlinespace[0.25em]
Generation & Pufibara   & 35/50   & 17.0  & 2.34 \\
           & Claude Code & 27/50   & 81.1  & 2.49 \\
\addlinespace[0.25em]
Tuning     & Pufibara   & 37/50   & 22.9  & 3.56 \\
           & Claude Code & 34/50   & 111.6 & 6.96 \\
\addlinespace[0.45em]
\midrule
\multicolumn{5}{@{}l}{\textit{(b) Claude Sonnet 5}} \\
\midrule
Workflow & Harness & Passed & Tokens (M) & Runtime (h) \\
\midrule
Repair     & Pufibara   & 131/132 & 36.2  & 3.68 \\
           & Claude Code & 125/132 & 177.7 & 8.76 \\
\addlinespace[0.25em]
Generation & Pufibara   & 32/50   & 13.6  & 2.73 \\
           & Claude Code & 26/50   & 77.1  & 2.91 \\
\addlinespace[0.25em]
Tuning     & Pufibara   & 39/50   & 20.5  & 3.17 \\
           & Claude Code & 36/50   & 86.7  & 6.15 \\
\midrule
\multicolumn{5}{@{}l}{\textit{(c) Pufibara relative to Claude Code}} \\
\midrule
Backend & Workflow & Pass-rate gain & Token reduction & Runtime reduction \\
\midrule
DeepSeek v4 Flash & Repair     & +4.5 pp  & 82.5\% & 58.4\% \\
                  & Generation & +16.0 pp & 79.0\% & 6.1\% \\
                  & Tuning     & +6.0 pp  & 79.5\% & 48.8\% \\
\midrule
Claude Sonnet 5   & Repair     & +4.5 pp  & 79.6\% & 58.1\% \\
                  & Generation & +12.0 pp & 82.4\% & 6.3\% \\
                  & Tuning     & +6.0 pp  & 76.4\% & 48.4\% \\
\bottomrule
\end{tabular*}
\end{table}

The observed advantage is not confined to one workflow, but it is largest in Model Generation. Relative to Claude Code, Pufibara gains 16 percentage points with DeepSeek v4 Flash and 12 points with Claude Sonnet 5 in Generation, compared with 4.5 points in Repair and six points in Tuning under each backend.

Under the repository-reported accounting, Pufibara has lower token totals and less sequential runtime in every comparison. Reported token reductions range from 76.4\% to 82.5\%. Runtime reductions range from 6.1\% to 58.4\%, with the smallest runtime difference occurring in Generation. 

\subsection{Beyond Executability}
\label{sec:beyond-executability}

Aggregate pass counts do not distinguish submissions that fail to compile or simulate from executable models with incorrect engineering behavior. Among the predefined 38 hard Model Generation tasks under Claude Sonnet 5, Claude Code produced executable final submissions that failed the benchmark-owned behavioral contract on 21 tasks, compared with four for Pufibara. These models produced valid simulation trajectories but did not satisfy the required behavior in the evaluated scenarios.

This distinction reflects the three evaluation stages used by the benchmark. Model checking determines whether a submitted model can be instantiated and checked. Simulation determines whether it can execute under the specified scenario. The behavioral contract then examines whether the resulting trajectories satisfy the task-specific engineering requirements. An execution-only evaluation would not detect the failures identified at this final stage.

The failure-stage contrast therefore shows that an important part of the observed difference occurs after executability has been established. It is consistent with maintaining engineering obligations across revisions and tying simulation evidence to the candidate that produced it. Because the evaluation compares complete harnesses, controlled ablations would still be needed to determine the contribution of individual mechanisms.

\section{Discussion}
\label{sec:discussion}

\subsection{Beyond Model Executability}

The evaluation supports a distinction between producing an executable physical system model and satisfying the engineering requirements that the model is intended to represent. Across two matched LLM backends and three workflow families, Pufibara has higher pass counts and lower reported resource use in all six comparisons. The consistent direction of these results does not establish universal superiority, but it indicates that the organization of the agent workflow remains consequential even when the underlying LLM, task set, Modelica environment, and evaluator are matched.

The largest pass-count difference occurs in Model Generation, which requires the agent to determine model structure, observable bindings, simulation strategy, behavioral adequacy, and what to submit. Repair provides comparatively direct model-checking or simulation feedback and operates near ceiling, while Tuning restricts action to parameters of a frozen model. This interpretation is consistent with the intended role of persistent engineering obligations and candidate-bound simulation evidence, but workflow differences do not establish a component-level causal explanation.

The analysis in \Cref{sec:beyond-executability} makes the failure boundary concrete. Among the predefined 38 hard Generation tasks under Claude Sonnet 5, executable but behaviorally incorrect final submissions occur on 21 tasks for Claude Code and four for Pufibara. Model checking establishes that a model can be elaborated, and successful simulation establishes that it can produce trajectories under a configured scenario. Neither alone establishes that those trajectories satisfy the intended engineering behavior. Simulation provides engineering evidence only when its observations are interpreted against the relevant obligation, scenario, and candidate.

\subsection{Implications for Agent Harnesses and Benchmarks}

Pufibara's contribution is not simulation feedback itself, which is standard in iterative engineering modeling. The architectural change makes the relationships among engineering obligations, observable bindings, simulation scenarios, candidate identity, evidence, and submission decisions explicit in the agent's execution protocol. A runnable artifact is therefore not an implicit endpoint. The agent must identify the exact candidate it intends to submit and relate that decision to candidate-bound evidence.

Instructions, context, or skills can encourage similar reasoning, but a harness preserves state across turns, binds evidence to candidate identity, and records revision and submission as auditable actions. Other systems can adopt this architecture. The transferable contribution is the harness pattern realized in Pufibara, not an exclusive reasoning strategy.

Engineering evidence remains task specific because models require different invariants, observables, scenarios, tolerances, and response conditions. The abstraction is not a fixed set of checks. It links obligations to observables and scenarios, evidence to identified candidates, and adjudication to revision or submission.

The benchmark complements this architecture by keeping official evaluation outside the agent loop. The agent maintains its working engineering state and selects the simulations used during development, but it does not author the contract that determines PASS or FAIL. Frozen benchmark-owned contracts evaluate the exact submitted artifact after revision has ended. This separation prevents self-defined validation from becoming the official success criterion and makes executable-but-behaviorally-incorrect submissions observable. The benchmark therefore evaluates an end-to-end agent workflow rather than only code generation or model executability.

Source-grounded task synthesis adds a complementary benchmark contribution. Executable reference models provide realistic physical structure, while new faults, engineering briefs, tuning targets, and private contracts define the exact tasks. The source model is therefore an engineering substrate rather than an answer that can be reproduced unchanged. This construction balances realism with reduced direct-reconstruction risk without claiming that public source components are absent from LLM pretraining.

\subsection{Limitations and Threats to Validity}

The evaluation compares Pufibara and Claude Code as complete harnesses across three Modelica workflow families under two LLM backends. Each task is run once for each harness-backend combination, so the results do not capture variation across repeated runs under the same condition. The complete-system comparison retains each harness's native workflow and request semantics and therefore does not isolate the contribution of individual harness mechanisms. Repeated trials and controlled ablations could address these limitations.

Public source models and library components may have appeared in LLM pretraining data. New faults, engineering briefs, tuning targets, and private evaluation contracts make unchanged reproduction insufficient to solve the tasks, but they cannot guarantee contamination-free evaluation. The Repair workflow focuses on model-checking and simulation failures, while explicit task-specific behavioral evaluation is provided by Generation and Tuning. A PASS result is limited to the fixed scenarios and acceptance contracts and does not imply formal verification or correctness outside the evaluated conditions \citep{sargent2013verification}.

Token use is normalized across harness-specific usage records and should not be interpreted as provider billing totals. Runtime is specific to the reported execution environment. The benchmark combines public, composed, and internally authored reference models, but its coverage of proprietary industrial workloads remains untested. The evaluation is also limited to Modelica, Claude Code as the comparison harness, and two LLM backends.
\section{Conclusion}
\label{sec:conclusion}

Physical system modeling requires more than producing models that compile and simulate. An agent must preserve engineering requirements across iterations, connect execution and simulation evidence to the candidate that produced it, and explicitly decide what to submit. This work presented Pufibara, an agent harness that makes these relationships persistent and auditable. It also proposed a source-grounded benchmark construction method and used it to build the Modelica Agent Workflow Benchmark for independently evaluating end-to-end agentic Modelica workflows.

Across 232 tasks and two matched LLM backends, Pufibara achieves higher pass counts than Claude Code for every workflow. With DeepSeek v4 Flash, Pufibara passes 202 tasks compared with 185 for Claude Code. With Claude Sonnet 5, Pufibara passes 202 tasks compared with 187 for Claude Code. Under the repository-reported token accounting, Pufibara records 76.4\%--82.5\% lower logical-token totals and 6.1\%--58.4\% less runtime.

These findings provide system-level evidence across the reported Modelica tasks and scenario-bounded evaluation contracts. They do not isolate the contribution of individual harness mechanisms or establish universal superiority. Future work can test the robustness of these results through repeated trials and controlled ablations, broaden the expert validation of evaluation contracts, and extend the approach to additional modeling languages and simulation environments. A particular priority is evaluating locally deployable LLMs. This is especially important for the Modelica ecosystem, where industrial models may contain proprietary designs, parameters, and engineering knowledge that must remain within controlled environments.

Pufibara operationalizes a central principle for physical system modeling agents: model executability is not a sufficient stopping criterion, and an agent should submit only when candidate-bound simulation evidence supports all applicable engineering obligations.
\section*{Artifact Availability}
\label{sec:artifact}

The public repositories contain documentation, summary results, and selected supporting files released with this paper. Most of the Pufibara implementation remains private. Complete benchmark tasks and evaluation files also remain private to reduce the risk of future benchmark contamination. Independent evaluation may later be supported through controlled research access or a hosted submission service that keeps hidden tasks and contracts confidential. Any such access would be governed by terms restricting redistribution and use for model training.

\begingroup
\hypersetup{urlcolor=black}
\bibliographystyle{unsrtnat}
\bibliography{references}
\endgroup

\appendix
\section{Agent Configurations}
\label{app:agent-configurations}

The evaluation compares Pufibara with Claude Code 2.1.226\footnote{\url{https://code.claude.com/docs/en/changelog}} using DeepSeek v4 Flash and Claude Sonnet 5. Within each backend condition, both harnesses use the same underlying model while retaining their native request semantics. Pufibara requests to DeepSeek v4 Flash use a temperature of 0.1, while Claude Code retains its native sampling semantics. DeepSeek v4 Flash does not expose a separate reasoning-effort setting. Claude Sonnet 5 is run without an explicit temperature setting and with medium effort. Each task is run once for each harness-backend combination. Every run starts in a fresh process, session, and workspace, with no cross-task reuse of conversation history, candidate artifacts, or engineering state. All tasks are executed sequentially.

\begin{table}[htbp]
\centering
\small
\renewcommand{\arraystretch}{1.15}
\caption{Per-task execution limits in the reported evaluation.}
\label{tab:workflow-limits}
\begin{tabular*}{\linewidth}{
    @{\extracolsep{\fill}}
    lcccc
    @{}
}
\toprule
Workflow
& Main wall-time limit
& Verification grace
& Max agent turns
& Max simulations \\
\midrule
Repair     & 900 s & 0 s   & 100 & --  \\
Generation & 900 s & 120 s & 100 & --  \\
Tuning & 900 s & 120 s & 100 & 100 \\
\bottomrule
\end{tabular*}

\vspace{2pt}
\begin{minipage}{\linewidth}
\footnotesize
\textit{Note:} Verification grace extends the main wall-time limit only when final verification initiated by an explicit submission is already in progress. It does not permit additional agent turns. One agent turn comprises one LLM response and its associated tool-call batch. A dash indicates that no separate simulation-count limit is imposed.
\end{minipage}
\end{table}

Token limits are nonbinding safety guards rather than target stopping criteria. Provider request retry is disabled. If an attempt fails before producing a valid capability measurement because of infrastructure, it is recorded as an infrastructure-invalid attempt and excluded. Any protocol-authorized replacement uses a fresh execution identity and the unchanged task configuration.

Both harnesses receive the same agent-visible task package and are scored against the same frozen evaluator contract. Pufibara exposes its workflow profiles and persistent engineering state, whereas Claude Code retains its native prompt, context, and tool-use semantics. Neither harness can inspect the private evaluator during the agent loop. The evaluator runs only on the exact explicit submission after that loop ends.

The Docker image \texttt{openmodelica/openmodelica:v1.26.1-minimal} was used for all Modelica checking and simulation. Reported runtime is the sum of valid sequential task wall times. The harness-native usage fields are mapped to the logical-token categories defined in \Cref{sec:experimental-setup} according to \Cref{tab:token-field-mapping}. When a runtime does not report a cache category separately, that category is not reconstructed.

\begin{table}[htbp]
\centering
\small
\caption{Mapping from harness-native usage fields to logical-token categories.}
\label{tab:token-field-mapping}

\begin{tabularx}{0.65\linewidth}{
    @{}
    >{\raggedright\arraybackslash}p{0.20\linewidth}
    >{\raggedright\arraybackslash}X
    @{}
}
\toprule
Logical category & Native usage fields \\
\midrule
Uncached input &
\texttt{input\_tokens}, \texttt{prompt\_tokens} \\
Cache creation &
\texttt{cache\_creation\_input\_tokens}, \texttt{cache\_write} \\
Cache read &
\texttt{cache\_read\_input\_tokens}, \texttt{cache\_read} \\
Output &
\texttt{output\_tokens}, \texttt{completion\_tokens} \\
\bottomrule
\end{tabularx}
\end{table}

\end{document}